\documentclass[conference]{IEEEtran}
\IEEEoverridecommandlockouts

\usepackage{amsmath,amssymb,amsfonts}
\usepackage{algorithmic}
\usepackage{graphicx}
\usepackage{mhchem}
\usepackage{textcomp}
\usepackage{hyperref}
\usepackage{xcolor}
\usepackage{multirow}
\usepackage{graphicx}
\usepackage{xcolor}
\usepackage{mathtools}
\usepackage{amsmath}
\usepackage{subcaption}
\usepackage{array}

\def\BibTeX{{\rm B\kern-.05em{\sc i\kern-.025em b}\kern-.08em
    T\kern-.1667em\lower.7ex\hbox{E}\kern-.125emX}}

\begin{document}

\title{Automated Security Assessment\\for the Internet of Things
\thanks{\IEEEauthorrefmark{5}Corresponding author.}
}

\author{\IEEEauthorblockN{Xuanyu Duan\IEEEauthorrefmark{1}, Mengmeng Ge\IEEEauthorrefmark{2}, Triet Huynh Minh Le\IEEEauthorrefmark{3}, Faheem Ullah\IEEEauthorrefmark{3},\\Shang Gao\IEEEauthorrefmark{1}, Xuequan Lu\IEEEauthorrefmark{1}\textsuperscript{,}\IEEEauthorrefmark{5} and M. Ali Babar\IEEEauthorrefmark{3}\textsuperscript{,}\IEEEauthorrefmark{4}}
\IEEEauthorblockA{\IEEEauthorrefmark{1}School of Information Technology, Deakin University, Geelong, Australia}
\IEEEauthorblockA{\IEEEauthorrefmark{2}School of Computing Technologies, RMIT University, Melbourne, Australia}
\IEEEauthorblockA{\IEEEauthorrefmark{3}School of Computer Science, The University of Adelaide, Adelaide, Australia}
\IEEEauthorblockA{\IEEEauthorrefmark{4}Cyber Security Cooperative Research Centre, Australia}
\IEEEauthorblockA{Emails: xuanyuduan@gmail.com, mengmeng.ge@rmit.edu.au, triet.h.le@adelaide.edu.au,\\ faheem.ullah@adelaide.edu.au, shang.gao@deakin.edu.au, xuequan.lu@deakin.edu.au, ali.babar@adelaide.edu.au}
}

\maketitle

\begin{abstract}
Internet of Things (IoT) based applications face an increasing number of potential security risks, which need to be systematically assessed and addressed. Expert-based manual assessment of IoT security is a predominant approach, which is usually inefficient. To address this problem, we propose an automated security assessment framework for IoT networks. Our framework first leverages machine learning and natural language processing to analyze vulnerability descriptions for predicting vulnerability metrics.
The predicted metrics are then input into a two-layered graphical security model, which consists of an attack graph at the upper layer to present the network connectivity and an attack tree for each node in the network at the bottom layer to depict the vulnerability information. This security model automatically assesses the security of the IoT network by capturing potential attack paths. We evaluate the viability of our approach using a proof-of-concept smart building system model which contains a variety of real-world IoT devices and potential vulnerabilities. Our evaluation of the proposed framework demonstrates its effectiveness in terms of automatically predicting the vulnerability metrics of new vulnerabilities with more than 90\% accuracy, on average, and identifying the most vulnerable attack paths within an IoT network. The produced assessment results can serve as a guideline for cybersecurity professionals to take further actions and mitigate risks in a timely manner.
\end{abstract}


\section{Introduction} \label{sec:intro}

Internet of Things (IoT) based applications are being widely deployed for providing intelligent services underpinning smart cyber-physical systems like smart buildings and smart citizens~\cite{taivalsaari2018development,madakam2015internet}. Nowadays, most of the efficiency and optimization efforts in various personal, social, and business sectors, such as healthcare, manufacturing, transportation, and smart home, are incorporating IoT based applications~\cite{gubbi2013internet}.
At the same time, IoT systems are also becoming a prime target for cybersecurity attacks~\cite{duc2017security}, due to the ever-growing amount of sensitive data that such systems support. Hence, there is an increasing realization that it is critical to timely identify the risks within the interconnected systems and deploy appropriate security countermeasures.

Researchers and practitioners have devised and applied a number of security assessment models to identify the security issues in IoT systems. A security assessment model analyzes an entity (e.g., IoT network) to make sure it complies with certain security objectives~\cite{leszczyna2018standards}. 
One type of security assessment models is graphical security models, that offer systematic ways for assessing security vulnerabilities of systems~\cite{Hong2012}. Attack graphs (AGs)~\cite{sheyner2002automated} and attack trees (ATs)~\cite{ingoldsby2010attack} are two of the most widely used ones. AGs analyze the security of networked systems by identifying all conceivable attack sequences used by attackers to reach the potential target areas of systems~\cite{enoch2019security}.
ATs describe the security of systems formally and systematically by presenting various means by which a system can be attacked~\cite{Hong2012}. However, these single-layered security models (e.g., AGs and ATs) do not scale well to an increasing number of devices in complex networks~\cite{hong2016towards}. To address such scalability issue, a Hierarchical Attack Representation Model (HARM)~\cite{Hong2012} was proposed. It combines AGs and ATs into a two-layer model where they can be constructed in parallel. HARM has been successfully applied to assess security of complex network, including IoT systems~\cite{ge2017framework}.

Despite the performance improvement, the security assessment using HARM is not fully automated~\cite{ge2017framework}.
In particular, the assessment metrics (i.e., the Common Vulnerability Scoring System (CVSS) severity scores~\cite{CVSS}) of vulnerabilities are not always available in vulnerability databases (e.g., National Vulnerability Database (NVD)~\cite{NVD}) as inputs for security assessment with HARM.\footnote{Vulnerability assessment is different from security assessment as the former is applied to individual devices and is a step in performing the security assessment of an interconnected system containing multiple devices.}
Some vulnerabilities only have vulnerability descriptions without the assessment metrics when being added into NVD~\cite{feutrill2018effect}, and these metrics are important for vulnerability assessment.
Some approaches have been proposed to automate software vulnerability assessment to address the issues in the prediction of severity scores based on a vulnerability description~\cite{le2019automated,spanos2018multi,yamamoto2015text,spanos2017assessment,le2021survey}. However, to the best of our knowledge, there has been no work on automated network-level security assessment with missing vulnerability-related metrics.

To fill this gap, our paper aims to design and implement a novel end-to-end security assessment framework for IoT networks.
Specifically, the framework consists of an automated vulnerability assessment model, a graphical security model and a visualization model.
The vulnerability assessment model uses Natural Language Processing (NLP) and Machine Learning (ML) techniques to process the vulnerability descriptions on NVD for predicting severity scores. The reason to adopt ML techniques rather than rule-based techniques is the emergence of new terms in the descriptions of new vulnerabilities which may cause the frequent update of rules with the inclusion of new terms. Taking the network connectivity information and vulnerability information of each node (including severity scores from the vulnerability assessment model) as inputs, the graphical security model produces potential attack paths and evaluates the security of the network using the attack path information and security metrics. Finally, the visualization model is used to visualize the attack paths with the highest security risks.

The main \textbf{contributions} of this paper are summarized as follows:
\begin{enumerate}
\item Propose an end-to-end security assessment framework combining automated vulnerability assessment and graphical security modeling to determine the attack probability, attack impact and risk in interconnected systems; 
\item Evaluate the framework using a smart building system;
\item Develop a visualization interface to generate network diagrams for better representation and understanding of security assessment results. 
\end{enumerate}

The rest of the paper is organized as follows. Section~\ref{sec:related_work} introduces the background knowledge and related work. Section~\ref{sec:proposed_framework} explains our proposed framework. Section~\ref{sec:evaluation} describes the implementation of the proposed framework as well as presents the evaluation and visualization results. Section~\ref{sec:discussion} discusses the findings and limitations of the proposed framework, followed by the conclusion and suggestions on future research directions in Section~\ref{sec:conslusion}.

\section{Related Work} \label{sec:related_work}

This section discusses current work on automated assessment of vulnerabilities and security assessment for IoT networks.

\subsection{Automated Assessment of Vulnerabilities} 

Approaches have been proposed for automated assessment of vulnerabilities. Guo et al.~\cite{guo2005automated} offered a featherweight virtual machine (FVM) solution to address the issue of vulnerability testing safety. The FVM technology enables a vulnerability assessment tool to test the exact duplicate of a production-mode network service while maintaining complete isolation of the production-mode network service from the testing process. Besides of ensuring safety, the vulnerability assessment support system presented in this work may also automate the entire vulnerability testing process, making it possible to conduct vulnerability testing automatically and frequently. Shah et al.~\cite{shah2014automated} developed NetNirikshak 1.0, an automated Vulnerability Assessment and Penetration Testing (VAPT) tool that assists organizations in assessing their applications/services and analyzing the security posture. This program identifies vulnerabilities in a target system's applications and services. The automatic report generated by the tool is emailed to a given email address, and all traces of the scan are deleted from the hard disk along with the report, ensuring the report's confidentiality. Le et al.~\cite{le2019automated} indicated that the method of automatically assessing software vulnerabilities based on NLP was affected by concept drift. Concept drift occurs as a result of lacking proper handling of new (out of vocabulary) terms in the vulnerability descriptions. Therefore, they proposed a vulnerability assessment method that incorporates both character and word features. Furthermore, they designed a time-based cross-validation method to identify the optimal ML model for predicting each vulnerability metric. Blinowski et al.~\cite{blinowski2020cve} used a vulnerability classification scheme for IoT devices based on real-world data. The authors first categorize the vulnerabilities into seven categories, then further classify them using standard descriptors in the Common Platform Enumeration (CPE). They leverage ML techniques to achieve automatic classification to mitigate the threats posed by new vulnerabilities.

\subsection{Security Assessment for IoT Systems}

The security of IoT systems has been explored by several studies. Radomirovic et al.~\cite{radomirovic2010towards} proposed an asynchronous communication network and a security model with fingerprint recognition capability based on fundamental assumptions and observations about potential security and privacy threats. This security model facilitates research on the security and privacy of protocols used in IoT networks. Ge et al.~\cite{ge2017framework} established a paradigm for modeling and assessing IoT security. The framework is used to develop a graphical security model and a security evaluator to automate the security analysis of IoT. The security evaluator assesses security using a variety of security metrics and outputs the analysis results via an analytic modeling and assessment tool called Symbolic Hierarchical Automated Reliability and Performance Evaluator (SHARPE) \cite{sahner2012performance}. Park et al.~\cite{park2017security} adopted an integrated fuzzy MCDM (FMCDM) technique to develop a framework for assessing the security of IoT services. The integrated method utilizes an analytic network process (ANP) combined with a decision-making trial and evaluation laboratory (DEMATEL) technique based on fuzzy set theory to raise the sensitivity of interrelationships amongst various security metrics. Wang et al.~\cite{wang2018iot} developed a security model based on blockchain and InterPlanetary File System (IPFS). The study and experimental findings demonstrate that many security issues associated with traditional IoT architecture can be avoided. In addition, system performance has been significantly improved on distributed large-capacity storage, concurrency and query.

Bugeja et al.~\cite{bugeja2019iotsm} designed an innovative IoT Security Model (IoTSM) based on the Software Assurance Maturity Model (SAMM) framework. It has been extended with new security practices and data collected from IoT practitioners. As a result, IoTSM can be used to organize strategies and discourse IoT security from an end-to-end perspective. Matheu-Garc{\'\i}a et al.~\cite{matheu2019risk} presented an IoT-specific security certification methodology that would enable various stakeholders to evaluate security solutions for large-scale IoT deployments in an automated manner. Additionally, it promotes customer openness at the IoT security level, as the methodology includes a label as one of the primary outcomes of the certification process. The certification methodology is an implementation of the ETSI-presented Risk-based Security Assessment and testing methodologies based on ISO 31000 and ISO 29119. It is built on top of a variety of security testing and risk assessment technologies and approaches targeting the IoT landscape. Martin et al.~\cite{martin2020towards} proposed a novel security model capable of analyzing the properties of IoT systems. The security model can assist designers of IoT systems in creating more secure systems by highlighting vulnerabilities and weakest links. Additionally, they intended to cover the data access requirements in future work and support the formalization and certification of access control mechanisms in IoT systems. Waraga et al.~\cite{waraga2020design} introduced an automated IoT security testbed for assessing the security of IoT devices. The proposed testbed is based on open source tools managed by an open source management system, where changes are provided to users via an easy-to-use GUI. This approach allows other researchers to use the modular structure of the testbed to create their own testing tools. This IoT testbed keeps track of all exploits and CVEs discovered for the devices under test, as well as the services it hosts through each port. This testbed could also generate reports automatically that contain the results of all devices.

\subsection{Summary}
Many prominent studies have been conducted on automating vulnerability assessment and utilizing security models to assess the security of IoT devices. However, few studies focused on combining these assessments to automate network assessment under missing vulnerability assessment metrics. To the best of our knowledge, our paper is the first to propose a model for automatically assessing the security of interconnected systems in terms of the probability, risk and impact of network attacks based on node/device vulnerabilities.

\section{Proposed Framework} \label{sec:proposed_framework}

\begin{figure}
\centerline{\includegraphics[scale=0.65]{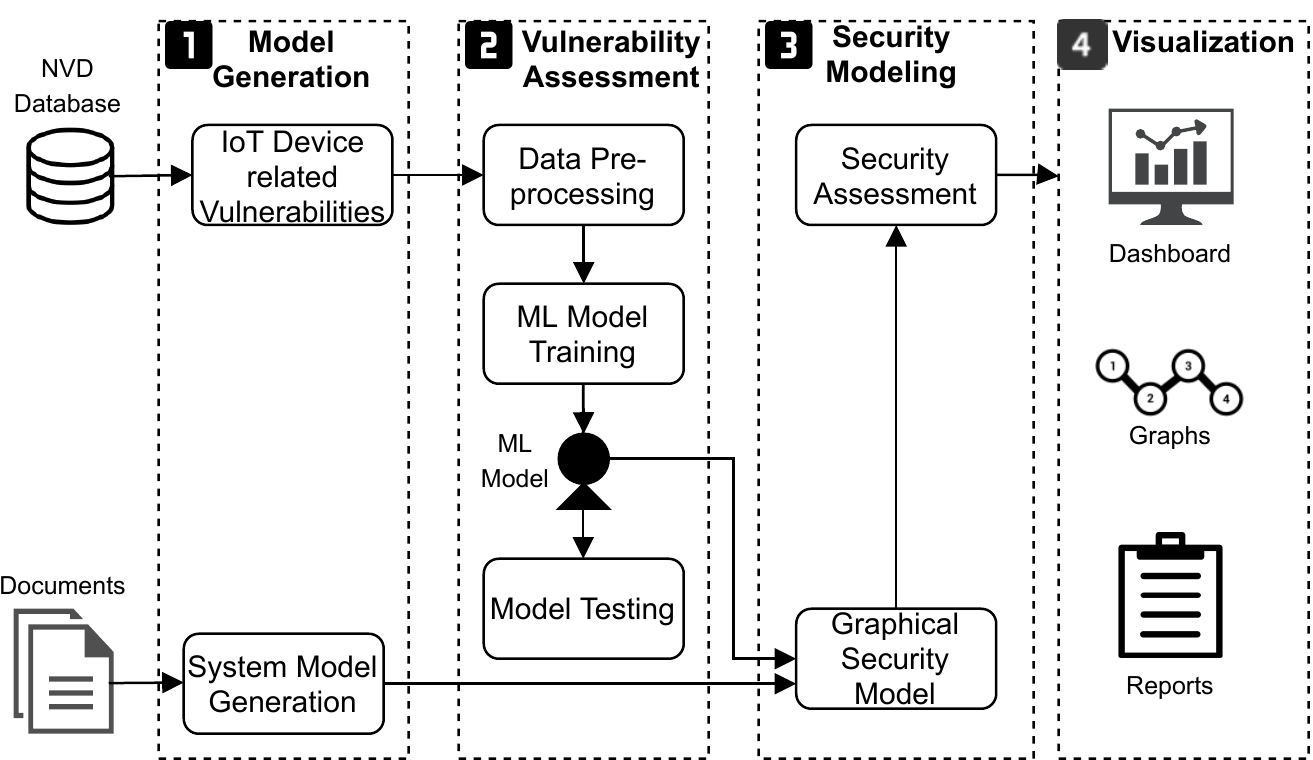}}
\caption{The proposed framework for automated security assessment}
\label{fig1}
\end{figure}

The proposed framework divides the security assessment process into four phases: system model generation, vulnerability assessment, graphical security model modeling and assessment, and data visualization, as illustrated in Fig.~\ref{fig1}. The implementation of the framework can be found at GitHub.\footnote{https://github.com/mmge88/automated-security-assessment} Each phase is discussed in the following subsections. 

\subsection{System Model Generation}

In \textbf{Phase 1}, we first consider a real-world IoT system and generate a system model based on the specifications of smart devices in the system and the connectivity of all devices. We then extract the vulnerability records from the NVD. Details of the vulnerability extraction can be found in Section~\ref{subsubsec: vul_filter}. This system model will later be combined with the graphical security model in \textbf{Phase 3} for the security assessment. The framework can be applied to any IoT system as long as the network connectivity and vulnerabilities of devices can be obtained.

\subsection{Vulnerability Assessment}

In \textbf{Phase 2}, we adopt the NLP techniques to preprocess the vulnerability descriptions and apply ML techniques to predict CVSS version 2 metrics based on the processed descriptions of vulnerabilities. We also utilize an over-sampling approach to mitigate the impact of imbalanced data. We tune the ML model to identify optimal hyperparameters and train the optimal model over the dataset downloaded from NVD after the preprocessing and over-sampling steps to predict the vulnerability scores of devices in the IoT system.

\subsection{Graphical Security Model and Assessment}

In \textbf{Phase 3}, we adopt a graphical security model based on a previous IoT network assessment research~\cite{ge2017framework} in which a two-layer HARM was utilized, with the upper layer (an attack graph for the network connectivity) and the lower layer (an attack tree for each node) to capture the information of how vulnerabilities can be exploited to gain a privilege.
The approach has been proven to be more efficient in identifying a network's most severe attack paths than single-layer attack graphs due to the parallel construction of the two layered models. As a result, we choose to retain this structure and adjust the program by integrating our automated vulnerability assessment module. The predicted scores from the vulnerability assessment are used as inputs to the graphical security model and compute the network-level security metrics to evaluate the security of the IoT system. The network-level security metrics are computed based on original scores of the vulnerabilities and the expected/ground-truth results are compared with prediction results.

\subsection{Data Visualization}

In \textbf{Phase 4}, we develop a web-based visualization interface to better present the assessment results. A network graph is used to visualize the relationships between all devices in the system and the attack path with highest risk based on the security evaluation results. A flask~\cite{grinberg2018flask} project is used to combine all network graphs and present them on the website.

\section{Evaluation} \label{sec:evaluation}
To evaluate the proposed framework, a smart building system is used as the test system where IoT technologies are heavily adopted.

\subsection{System Model Generation} \label{sec:system_model_generation}

The design of the test system is mainly derived from~\cite{hachem2020modeling,jia2019adopting}. Hachem et al.~\cite{hachem2020modeling} conducted a study on the smart building of the Adelaide University Health and Medical School (AHMS). Jia et al.~\cite{jia2019adopting} undertook an assessment on the adoption of IoT in smart building applications and categorized them according to their critical functions and aims, which improved our test system based on the AHMS. The test system includes major functionalities of a smart building system that leverages IoT techniques.

\begin{figure*}
     \centering
     \begin{subfigure}[b]{0.24\textwidth}
         \centering
         \includegraphics[width=\linewidth]{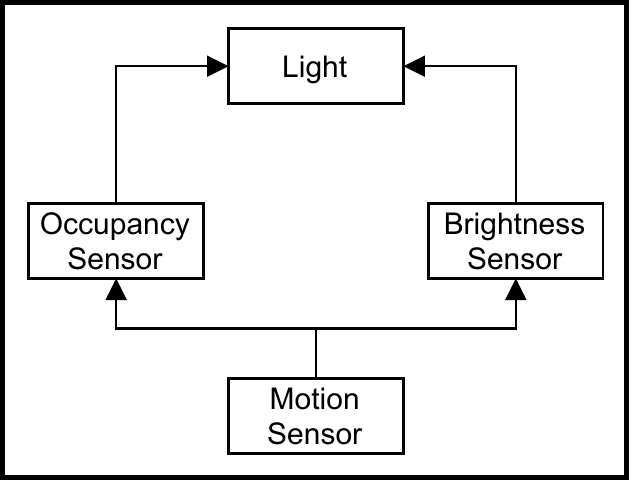}
         \caption{Lighting}
         \label{fig:lighting_system}
     \end{subfigure}
     \hfill
     \begin{subfigure}[b]{0.24\linewidth}
         \centering
         \includegraphics[width=\linewidth]{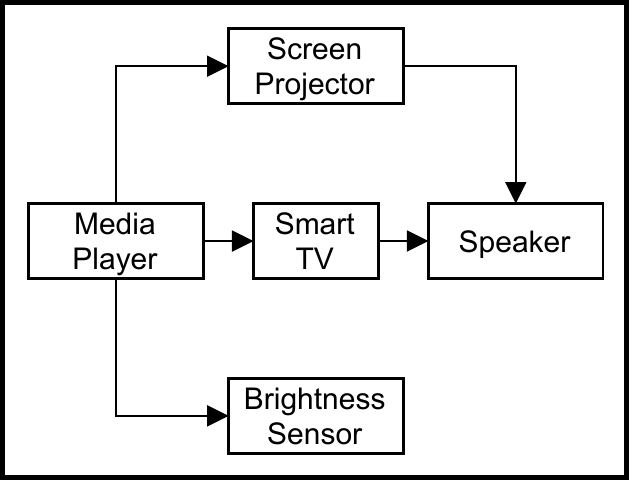}
         \caption{Audiovisual}
         \label{fig:audiovisual_system}
     \end{subfigure}
     \begin{subfigure}[b]{0.24\linewidth}
         \centering
         \includegraphics[width=\linewidth]{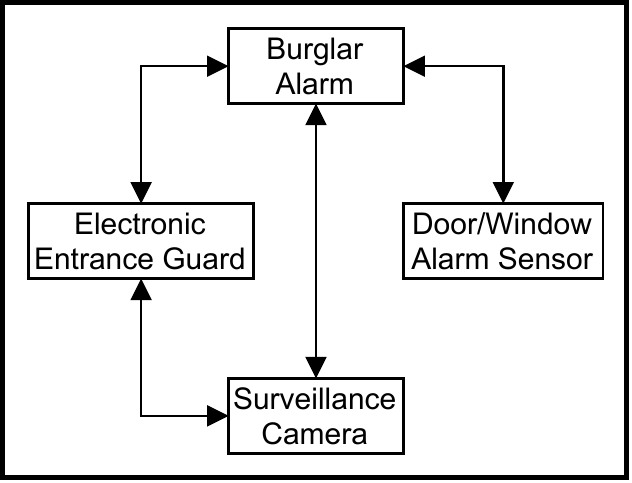}
         \caption{Security}
         \label{fig:security_system}
     \end{subfigure}
    \hfill
     \begin{subfigure}[b]{0.24\linewidth}
         \centering
         \includegraphics[width=\linewidth]{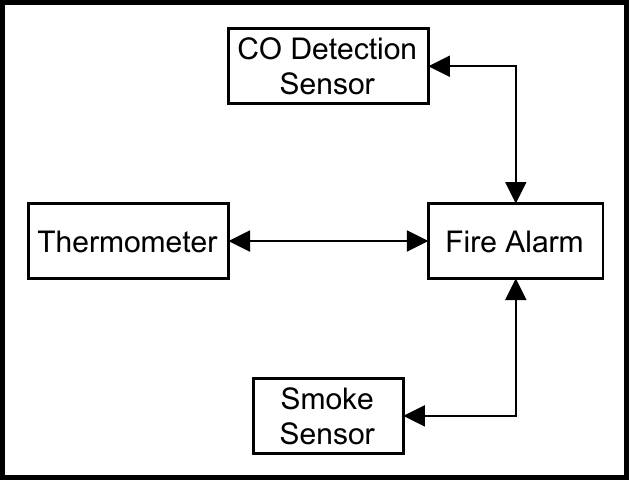}
         \caption{Fire detection}
         \label{fig:fire_detection_system}
     \end{subfigure}
     \begin{subfigure}[b]{0.245\textwidth}
         \centering
         \includegraphics[width=\textwidth]{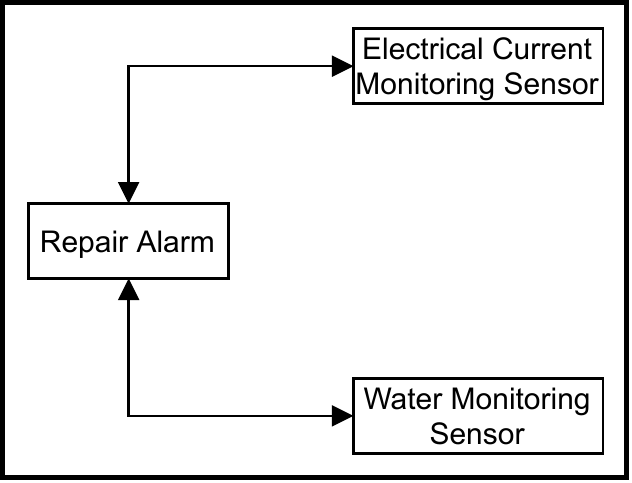}
         \caption{Maintenance}
         \label{fig:maintenance_system}
     \end{subfigure}
     \begin{subfigure}[b]{0.245\textwidth}
         \centering
         \includegraphics[width=\textwidth]{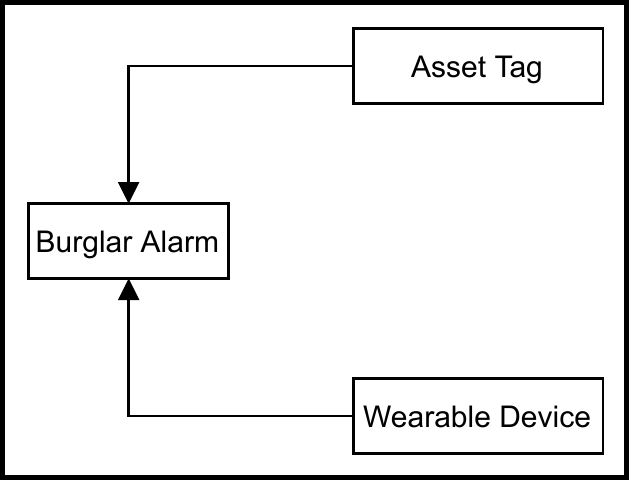}
         \caption{Resource tracking}
         \label{fig:resource_tracking_system}
     \end{subfigure}
     \begin{subfigure}[b]{0.49\textwidth}
         \centering
         \includegraphics[width=\textwidth]{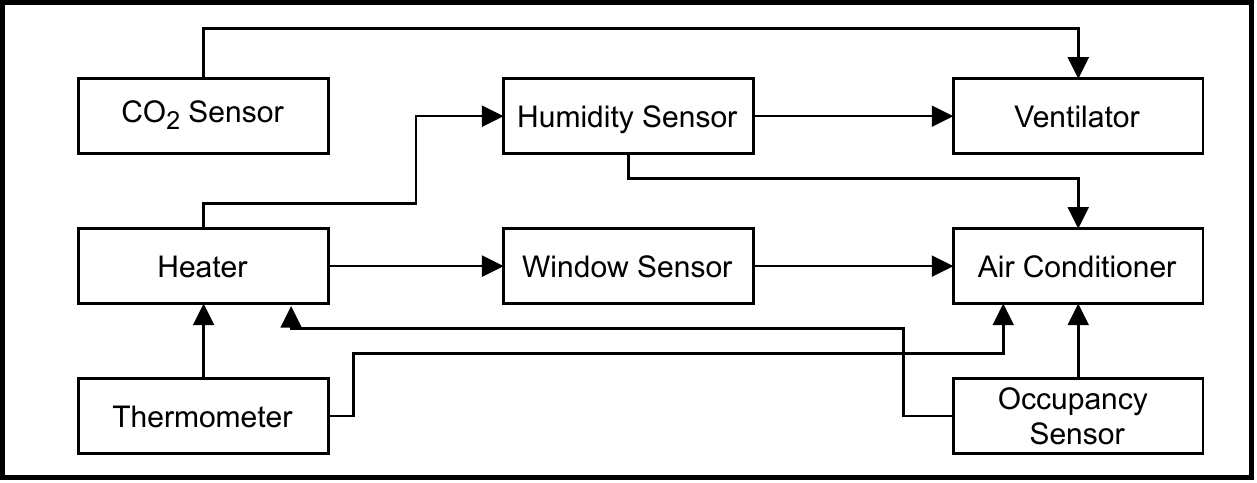}
         \caption{HVAC}
         \label{fig:HVAC_system}
     \end{subfigure}
    
        \caption{Smart building subsystems}        
        \label{fig:subsystem_model}
\end{figure*}

\begin{figure}
     \centering
     \begin{subfigure}[t]{0.48\textwidth}
         \centering
         \includegraphics[width=\textwidth]{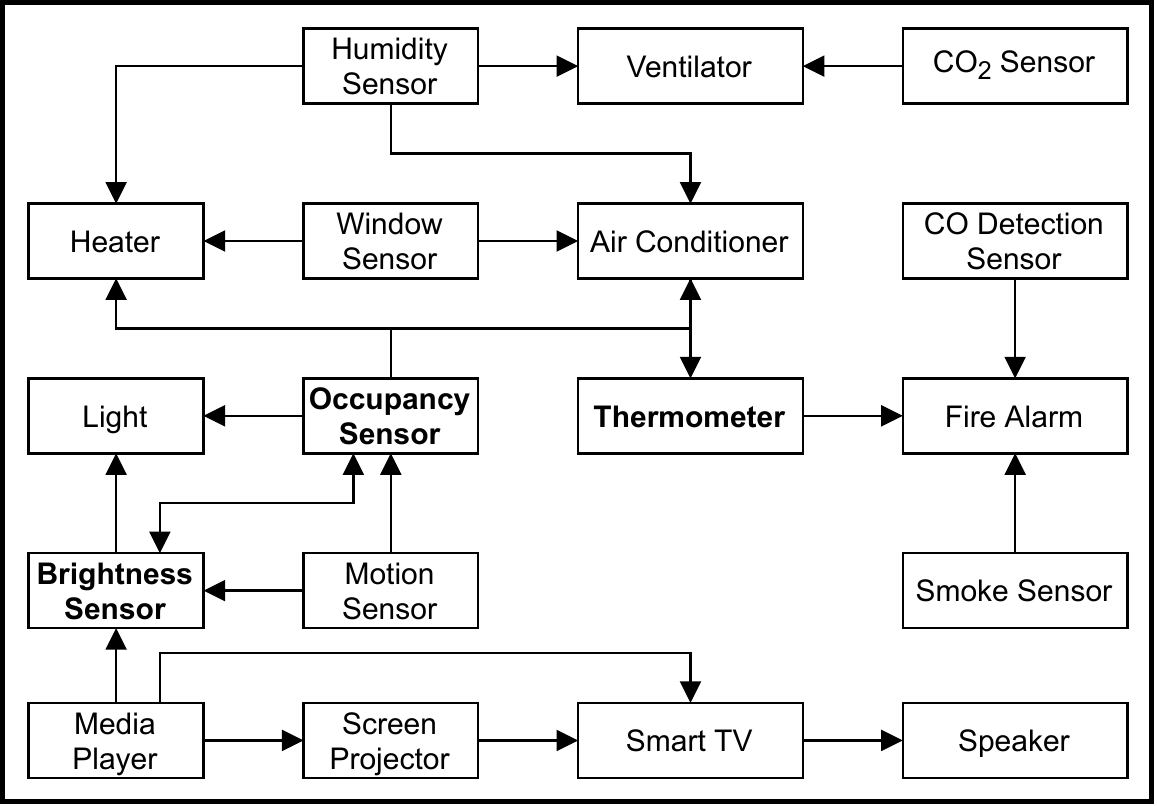}
         \caption{Combined smart building automation system}
         \label{fig:combined_smart_building_automation_system}
     \end{subfigure}
       \par\bigskip
     \begin{subfigure}[t]{0.48\textwidth}
         \centering
         \includegraphics[width=\textwidth]{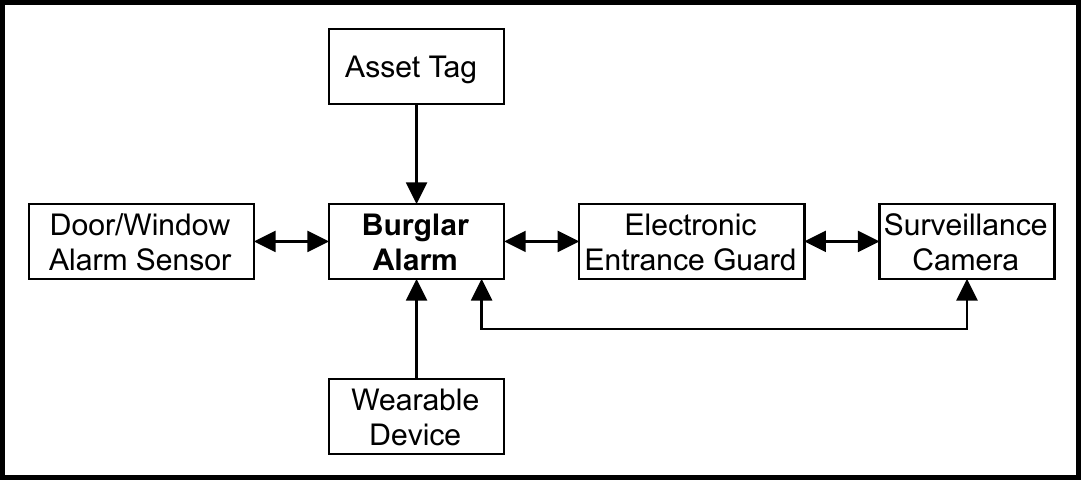}
         \caption{Combined security system}
         \label{fig:combined_security_system}
     \end{subfigure}
        \caption{Combined systems in the smart building system}        
        \label{fig:combined_system_model}
\end{figure}

Our test system incorporates a variety of smart sensors and consists of seven subsystems, each of which serves a distinct purpose and is in charge of implementing a part of functionalities in the smart building.
The seven subsystems are: lighting, audiovisual, security, fire detection, maintenance, resource tracking and HVAC subsystems, as shown in Fig.~\ref{fig:subsystem_model}. Since several subsystems have shared sensors, we obtain two combined subsystems by linking them via the common devices. As shown in Fig.~\ref{fig:combined_system_model}, the security and the resource tracking are combined via the burglar alarm device, named as a combined security system. The audiovisual, lighting, fire detection and HVAC systems are combined via the brightness sensor, occupancy sensor and thermometer, referred to as a combined smart building automation system. The maintenance system does not have a common device with other subsystems thus not being included in any combined system.

\subsubsection{Design and generation of subsystems} The devices and functionality of each subsystem are described in Table~\ref{tab:system}. As devices of the same type are usually purchased from the same manufacturer (e.g., light sensors in the lighting subsystem), we assume they have identical vulnerabilities. We select one device from each type in the test system to capture attack scenarios among different types of devices. More device types can be added and modeled via the framework.

\begin{table*}
\caption{Devices and functionality of the subsystems}
\label{tab:system}
\begin{tabular}{|m{2.4cm}|m{6.0cm}|m{8.5cm}|}
\hline
\textbf{Subsystem} & \textbf{Devices} & \textbf{Functionality}\\
\hline
Lighting & 2 light sensors, 1 light controller, 1 occupancy sensor, 1 brightness sensor, 1 motion sensor & Illuminate the smart building by controlling the lights in response to occupancy and the brightness sensors in response to the motion sensor \\
\hline
Audiovisual & 1 media player, 1 screen projector, 1 smart TV, 1 brightness sensor, 1 speaker &  Regulate the operation of audiovisual devices in response to users\\
\hline
Security & 2 door and window alarm sensors, 1 burglar alarm, 1 electronic entrance guard, 1 surveillance camera & Monitor the security of the smart building and respond to security incidents \\
\hline
Fire detection & 1 CO detection sensor, 1 thermometer, 1 smoke sensor, 1 fire alarm & Detect and prevent potential fire incidents based on the carbon monoxide concentration, temperature and smoke concentration \\
\hline
Maintenance & 1 electrical current monitoring sensor, 1 water monitoring sensor, 1 repair alarm & Monitor and record the usage of electricity and water and report problems in a timely manner \\
\hline
Resource tracking & 2 asset tags, 2 wearable devices, 1 burglar alarm & Monitor and track the status of assets and wearable devices, and protect these resources from being stolen or damaged \\
\hline
HVAC & 1 heater, 1 air conditioner, 1 ventilator, 1 humidity sensor, 1 \ce{CO2} sensor, 1 thermometer, 2 window sensors, 1 occupancy sensor & Increase the comfort level of the smart building by controlling the temperature and humidity \\
\hline
\end{tabular}
\label{tab:subsystem}
\end{table*}

\subsubsection{Vulnerability filtering} \label{subsubsec: vul_filter}

To filter vulnerabilities for the related devices, the vulnerability data dated from 2002 to 2020 is first downloaded from NVD, which contains 144,345 vulnerability records. All of the records are merged and potential vendors of IoT devices are identified. The records associated with the potential vendors are extracted. Then, the records related to the subsystems are extracted after we manually examine whether the vulnerability descriptions are associated with the corresponding subsystems. Subsequently, an extraction based on the potential subsystem devices is performed. This set of extracted vulnerability data will be used for model training, validation and testing in the vulnerability assessment process. The dataset is split into 80\% for training, 10\% for validation, and 10\% for testing, excluding the evaluation set.

Finally, the vulnerability records related to the actual subsystem devices of various vendors (e.g., Broadcom, Panasonic, and Schneider Electric) are extracted. We utilize these device related vulnerabilities as evaluation data in the vulnerability assessment process and take the predicted results as inputs into the graphical security modeling and assessment. The number of records used in the vulnerability assessment is shown in Table~\ref{tab:dataset}. Note that the vulnerabilities in each dataset are unique and do not overlap with the other sets.

\begin{table}
\caption{The number of unique vulnerabilities in each dataset used in our study}
\begin{center}
\centering
\begin{tabular}{|m{2.0cm}|m{4.0cm}|}
\hline
\textbf{Dataset} & \textbf{No. of vulnerabilities/records}\\
\hline
Training & 115,395 \\
\hline
Validation & 14,424 \\
\hline
Testing & 14,424 \\
\hline
Evaluation & 102 \\
\hline
\end{tabular}
\label{tab:dataset}
\end{center}
\end{table}

\subsection{Vulnerability Assessment}

\subsubsection{Data pre-processing}
Each vulnerability record collected from NVD consists of a description, CVSS v2 metrics (i.e., base score metric group captures the characteristics of a vulnerability that do not change with time and across user environment), and privileges. In particular, the base score is measured by impact and exploitability. The impact is based on confidentiality, integrity, and availability. The exploitability is identified by the access vector, access complexity and authentication. There are three types of privileges, including all (i.e., root), user, and other privileges~\cite{scoringsystem}. 

To preprocess these records, the descriptions are first passed through NLP. For each vulnerability record, the stop words are removed, and all words in the description are converted to lowercase. Then term frequency-inverse document frequency (TF-IDF) vectorizer with n-gram (1-3) is applied to transform the text to vectors. TF-IDF is adopted because it has been proved to be well performed for vulnerability analysis \cite{le2019automated}. Other relevant algorithms are to be investigated in our future work. The processed vulnerability descriptions are taken as the input to the ML model. The CVSS metrics and privilege values are predicted as the output.

\begin{figure}
     \centering
     \begin{subfigure}[t]{0.48\textwidth}
         \centering
         \includegraphics[width=\textwidth]{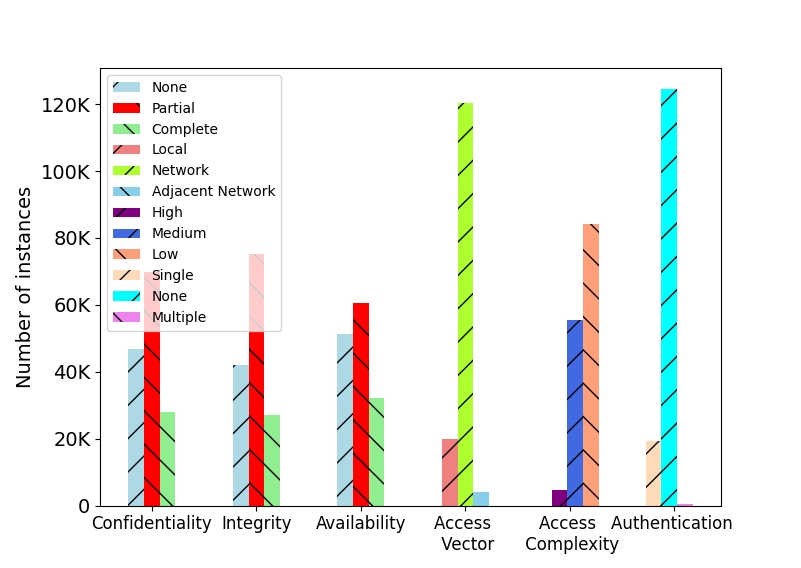}
         \caption{Number of label instances for each metric in the base score}
         \label{fig:frequencies_of_base_score}
     \end{subfigure}
    
     \begin{subfigure}[t]{0.48\textwidth}
         \centering
         \includegraphics[width=\textwidth]{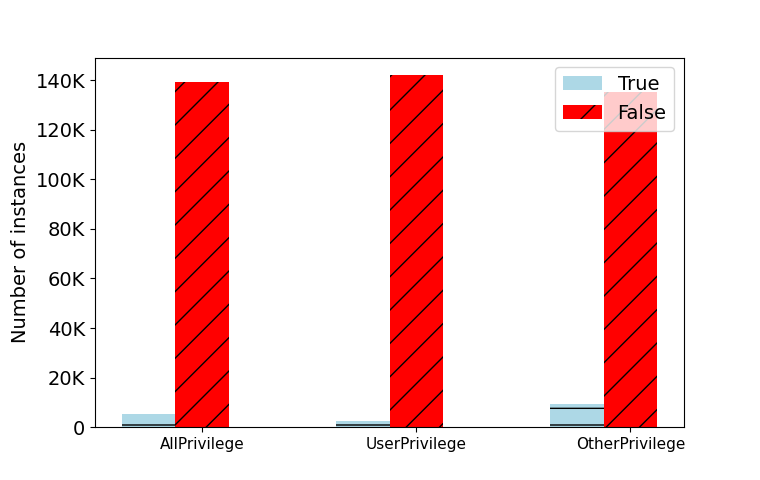}
         \caption{Number of label instances for privileges}
         \label{fig:frequencies_of_labels_privilege}
     \end{subfigure}
        \caption{Number of label instances before sampling}        
        \label{fig:frequencies_of_labels}
\end{figure}

\begin{figure}
     \centering
     \begin{subfigure}[t]{0.48\textwidth}
         \centering
         \includegraphics[width=\textwidth]{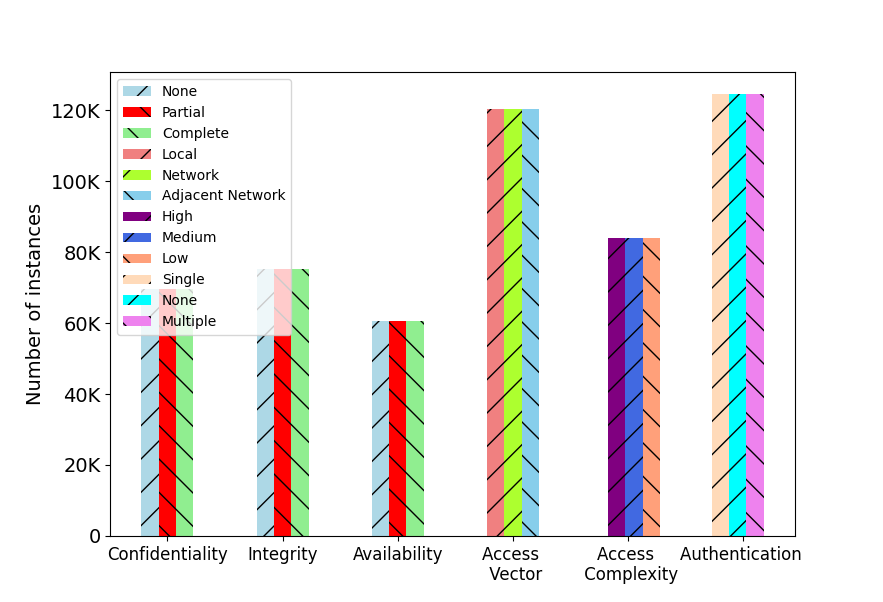}
         \caption{Number of label instances for each metric in the base score.}
         \label{fig:oversampling_of_base_score}
     \end{subfigure}
   
     \begin{subfigure}[t]{0.48\textwidth}
         \centering
         \includegraphics[width=\textwidth]{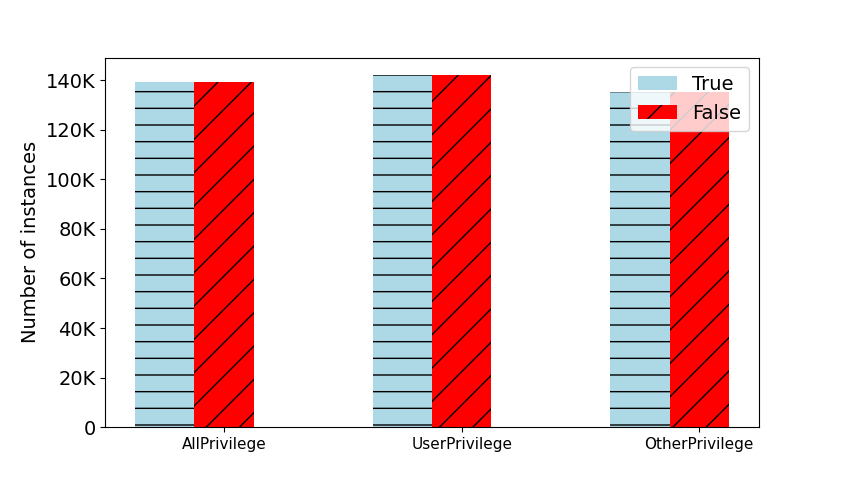}
         \caption{Number of label instances for privileges }
         \label{fig:oversampling_privilege}
     \end{subfigure}
        \caption{Number of label instances after sampling}        
        \label{fig:oversampling}
\end{figure}

We evaluate the processed data before predicting the six metrics in CVSS base score and types of privileges. As illustrated in Fig.~\ref{fig:frequencies_of_base_score}, each of the six metrics has three distinct values. A multi-class classification is needed from an ML perspective. In comparison, Fig.~\ref{fig:frequencies_of_labels_privilege} shows two values for each privilege, addressing the binary classification in ML. Additionally, as indicated in both figures, the distribution of each class is severely imbalanced for access vector, access complexity, authentication, and the three types of privileges. To mitigate the impact of imbalanced data and avoid the ignorance of critical data during model training, we utilize an over-sampling approach, RandomOverSampler~\cite{JMLR:v18:16-365}, to duplicate the instances in the minority class. By randomly reproducing instances in a minority class, this approach raises the number of instances until the minority and majority classes are balanced. The dataset after oversampling is shown in Fig.~\ref{fig:oversampling}.

\subsubsection{Model tuning}
The model tuning is done using the training and validation datasets, as shown in Table~\ref{tab:dataset}. Specifically, for tuning, a model is first trained on the training set and then evaluated on the validation set.  The Light Gradient Boosting Machine (LGBM)~\cite{LightGBM} is used for the classification. LGBM has been proven to outperform other ML models (such as Naive Bayes, Logistic Regression, Support Vector Machine, Random Forest and XGBoost~\cite{chen2016xgboost}) and thus being the optimal model for classifying vulnerability assessment metrics~\cite{le2019automated}. Its primary advantage is scalability, as the sub-trees are formed leaf-wise rather than level-wise compared to other gradient boosting techniques \cite{LightGBM}. As indicated previously, the prediction of metrics in the base score is a multi-class classification problem, whereas the prediction of privileges is a binary classification problem. As such, we set the objective for predicting metrics of the base score to multi-class and privileges to binary. The complexity of LGBM is mainly determined by the maximum depth and leaf count~\cite{LightGBM}. Therefore, we use 100, 300 and 500 as the number of leaves for tuning. To limit the tree depth and avoid overfitting~\cite{LightGBM}, we set 100, 200 and 300 as the maximum depth during tuning. These values follow the previous practices in this domain (e.g.,~\cite{le2019automated,le2020puminer,le2021deepcva}).

Additionally, the grid search approach is adopted to determine the optimal (with the highest value of evaluation metric on the validation set) combination of hyperparameters by considering all possible hyperparameter combinations. We set the evaluation metric as F1 score to tune the model for the binary classification. F1-score is a measure of predictive effectiveness and capable of balancing recall (i.e., the number of correct positive class predictions among all actual positive instances) and precision (i.e., the number of correct positive class predictions among all positive predictions)~\cite{scikit-learn}. 

Due to the fact that the prediction of metrics in the base score is a multi-class classification problem, we quantify the result using the F1-weighted method, i.e. calculate F1 score for each label and use the number of true instances for each label as the weight for the average value.
The model with the highest F1-weighted score on the validation set is selected as the optimal model.
Table~\ref{tab1} illustrates the optimal combination of hyperparameters for each privilege where ``no. leaves" denotes number of leaves, ``max depth" denotes maximum depth, ``std\_dev" represents the standard deviation and the time unit of ``running time" is second. Table~\ref{tab2} illustrates the optimal combination of hyperparameters for each metric. We will perform sensitivity analysis to examine the impact of different combinations of hyperparameters on the model performance in the future work.

\begin{table}
\caption{The validation results (F1 score) and the optimal combination of hyperparameters (i.e., max depth and number of leaves) for each privilege}
\begin{center}
\centering
\begin{tabular}{|m{1.7cm}|m{0.6cm}|m{0.55cm}|m{0.6cm}|m{0.8cm}|m{0.6cm}|m{1.0cm}|}
\hline
\textbf{Privilege} & \textbf{Max depth} & \textbf{No. leaves} & \textbf{Mean} & \textbf{Std\_dev}  & \textbf{F1 score} & \textbf{Running time (s)}\\
\hline
All privilege & 300 & 500 & 0.942 & 0.003 & 0.978 & 133.615 \\
\hline
Other privilege & 300 & 500 & 0.914 & 0.006 & 0.965 & 130.612 \\
\hline
User privilege & 300 & 500 & 0.964 & 0.003 & 0.988 & 132.743 \\
\hline
\end{tabular}
\label{tab1}
\end{center}
\end{table}

\begin{table}
\caption{The validation results (F1-weighted score) and the optimal combination of hyperparameters (i.e., max depth and number of leaves) for each CVSS metric in base score}
\begin{center}
\begin{tabular}{|m{1.55cm}|m{0.55cm}|m{0.6cm}|m{0.55cm}|m{0.8cm}|m{0.9cm}|m{0.95cm}|}
\hline
\textbf{Metric} & \textbf{Max\newline depth} & \textbf{No.\newline leaves} & \textbf{Mean} & \textbf{Std\_dev} & \textbf{F1-weighted} & \textbf{Running\newline time (s)}\\
\hline
Confidentiality & 200 & 500 & 0.825 & 0.007 & 0.895 & 394.004\\
\hline
Integrity & 100 & 500 & 0.848	& 0.004 & 0.914 & 407.055\\
\hline
Availability & 100 & 100 & 0.806 & 0.003 & 0.835 & 317.599\\
\hline
Access vector & 200 & 500 & 0.953 & 0.003 & 0.988 & 461.245\\
\hline
Access complexity & 300 & 500 & 0.825 & 0.005 & 0.913 & 401.231\\
\hline
Authentication & 100 & 300 & 0.944 & 0.002 & 0.974 & 389.540\\
\hline
\end{tabular}
\label{tab2}
\end{center}
\end{table}

\subsubsection{Model testing}
The optimal hyperparameters from the previous phase are used to build the model for testing. The model is trained with the training and validation datasets and is then tested on the testing set, as given in Table~\ref{tab:dataset}.
The F1 score and accuracy score are utilized to evaluate the model performance for the binary classification problem. F1-weighted, F1-macro and accuracy are used for the multi-class problem. The F1-macro calculates the F1 score of each label and finds an unweighted mean. 

As shown in Table~\ref{tab3} and Table~\ref{tab4}, all scores for the prediction of privileges are greater than 95\%, and scores for the prediction of integrity and exploitability metrics achieve 90\%. Scores for predicting confidentiality are approximately 90\%, while scores for predicting availability are roughly 83\%. 
This indicates that the model performs well with the hyperparameters tuned in the previous stage.

\begin{table}
\caption{Test results for privileges}
\begin{center}
\begin{tabular}{|c|c|c|}
\hline
\textbf{Privilege} & \textbf{Accuracy Score} & \textbf{F1 Score} \\
\hline
All privilege & 0.9786 & 0.9789  \\
\hline
Other privilege & 0.9680 & 0.9693 \\
\hline
User privilege & 0.9878 & 0.9879 \\
\hline
\end{tabular}
\label{tab3}
\end{center}
\end{table}

\begin{table}
\caption{Test result for metrics in the base score}
\begin{center}
\begin{tabular}{|m{2.2cm}|m{1cm}|m{0.8cm}|m{1cm}|}
\hline
\textbf{Metric} & \textbf{Accuracy Score} & \textbf{F1-macro} & \textbf{F1-weighted} \\
\hline
Confidentiality & 0.8910 & 0.8908 & 0.8905 \\
\hline
Integrity & 0.9145 & 0.9138 & 0.9140	\\
\hline
Availability & 0.8358 & 0.8347 & 0.8344 \\
\hline
Access vector & 0.9886 & 0.9886 & 0.9886 \\
\hline
Access complexity & 0.9126 & 0.9124 & 0.9122 \\
\hline
Authentication & 0.9741 & 0.9741 & 0.9741 \\
\hline
\end{tabular}
\label{tab4}
\end{center}
\end{table}

\subsubsection{Severity score prediction}
With the trained model, we use the evaluation data (extracted vulnerability data related to the subsystem devices, as described in Section~\ref{subsubsec: vul_filter}) to predict metrics values in base score and privileges. The predicted exploitability and impact metrics are then utilized to calculate the predicted CVSS base/severity score using the CVSS equations \cite{scoringsystem}, as given in equations~\eqref{eq1},~\eqref{eq2},~\eqref{eq3} and~\eqref{eq4}. The ConfImpact, IntegImpact and AvailImpact quantify the impact of a successfully exploited vulnerability on the confidentiality, integrity and availability of a system of interest, respectively.
The $f(\text{Impact})$ depends on the value of Impact. More details can be found in~\cite{scoringsystem}.

\begin{equation}
\fontsize{8}{9}\selectfont
\begin{aligned}
\label{eq1}
 \text{Base score} = (.6 \times \text{Impact} +.4 \times \text{Exploitability} - 1.5) \times f(\text{Impact})
 \end{aligned}
\end{equation}
\begin{equation}
\fontsize{8}{9}\selectfont
\label{eq2}
 \begin{aligned}
 \text{Impact} =~& 10.41 \times (1 - (1 - \text{ConfImpact}) \times (1 -  \text{IntegImpact})~\times\\ & (1 - \text{AvailImpact}))
 \end{aligned}
\end{equation}
\begin{equation}
\fontsize{8}{9}\selectfont
\begin{aligned}
\label{eq3}
 \text{Exploitability} = 20 \times \text{AccessComp.} \times \text{Authentication} \times \text{AccessVector}
 \end{aligned}
\end{equation}
\begin{equation}
\fontsize{8}{9}\selectfont
\label{eq4}
f(\text{Impact}) = \Bigg\{
	\begin{aligned}
		0, \text{if Impact} = 0  \\
		1.176, \text{otherwise}
	\end{aligned}
\end{equation}

\subsection{Graphical Security Model and Assessment}

To better illustrate the assessment results in the network level, a graphical security model is adopted. It is built on a previous Python project \cite{ge2017framework} with modified modules to accommodate the requirements of this work. 

\begin{figure}
\centerline{\includegraphics[scale=0.5]{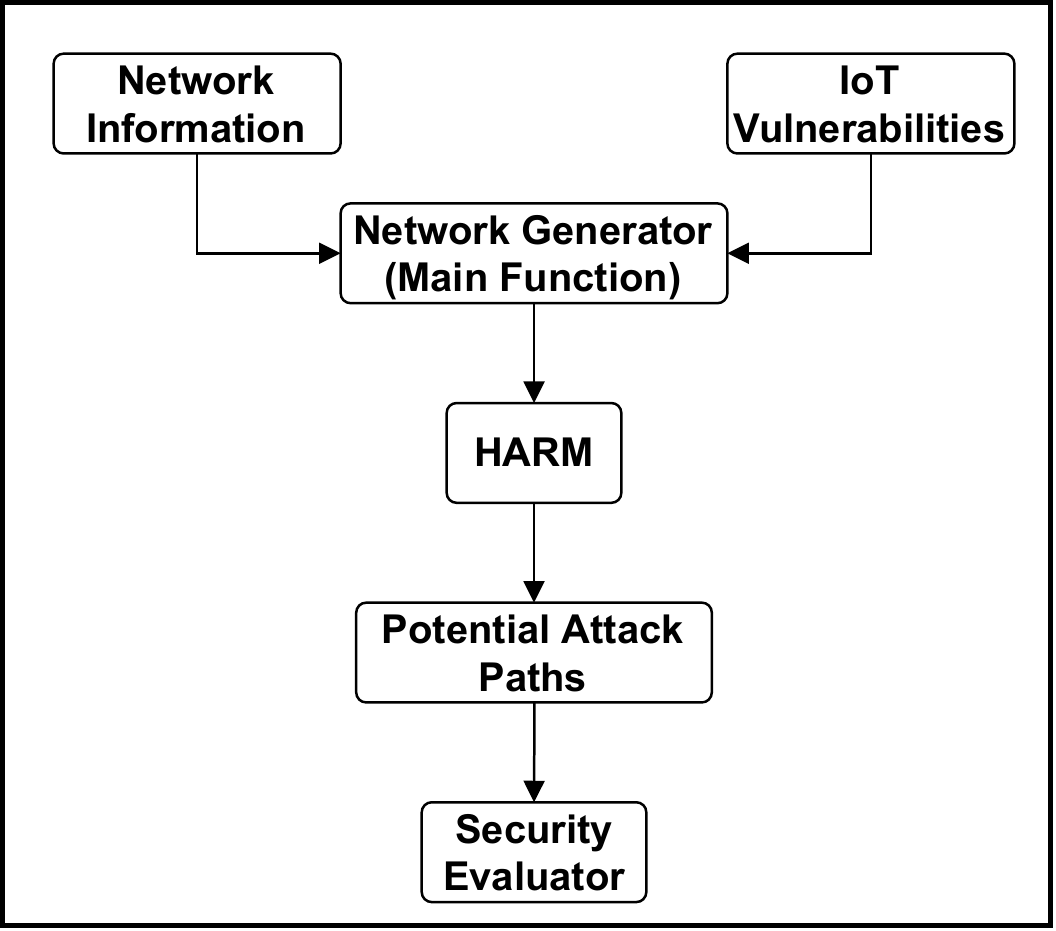}}
\caption{Structure of the Graphical Security Model}
\label{fig:code_structure}
\end{figure}

\subsubsection{Model description} As illustrated in Fig.~\ref{fig:code_structure}, the network generator contains the main function for constructing a smart building network of devices and vulnerabilities in the IoT context. It first initializes each device (i.e., node in the program) with the prediction results from the vulnerability assessment by representing the exploitation of vulnerabilities on each device based on privileges and assigning impact and exploitability scores to each vulnerability for network-level metric calculation. In particular, an integer value from 1 to 3 is assigned to each Privilege (All/Other/User), representing the access level an attacker obtains when a node is compromised. When ``other privilege" is true, the default privilege value of 1 is retained; the value rises to 2 if ``user privilege" is true, and to 3 if ``all privilege" is true. We primarily focus on the three security metrics for the security assessment of networks~\cite{CVSS}: the probability of a successful attack, the impact of an attack, and the risk of an attack. The probability of a successful attack (ranging from 0 to 1) is measured by the exploitability score (in equation~\eqref{eq3} ranging from 0 to 10) divided by 10, the impact of an attack refers to impact score (in equation~\eqref{eq2}), and the risk of attack is impact times probability. The network generator then connects the devices according to the pre-defined topology of each subsystem presented in Section \ref{sec:system_model_generation}.

After that, the generated network is taken as input to the graphical security model generator, generating a Hierarchical Attack Representation Model (HARM)~\cite{Hong2012}. This HARM consists of an upper layer attack graph and a lower layer attack tree for each vulnerable node. To calculate the attack paths, we also specify the potential entry points and attack targets in the network. Using the lighting subsystem as an example, the sensors can be used as entry points due to their lack of security protections; the light controller can be used as the attack target as the attacker can use the compromised controller to control all lights.

The most significant attack paths (e.g., with the maximum risk) can be identified through the security metrics computed via the security evaluator. We use a bottom-up approach to calculate the metrics via the HARM. Values in an attack tree are calculated first and populated to the node via the security evaluator. Multiple exploitable vulnerabilities may exist on a single device, which can be analyzed by two types of gates in the attack tree: $AND$ gate and $OR$ gate. An $AND$ gate means that all vulnerabilities must be exploited to compromise a node. In contrast, an $OR$ gate implies that an attacker can gain control of the node by exploiting only one of the vulnerabilities. These two scenarios result in different evaluation methods in the security evaluator, where $AND$ is accumulative and $OR$ is selective. In the attack path level, we calculate the metric value by accumulating the values of all nodes along the path (i.e., multiplication of probability values of nodes; sum of impact/risk values of nodes). We then choose the attack path with the highest risk value in the network-level assessment. Security evaluation results are passed back to the main function for final output and visualization. When comparing the ground-truth and predicted values at the network level, we use the percentage error to determine the model's prediction accuracy by Equation~\ref{eq5}. 

\begin{equation}
\label{eq5}
\begin{aligned}
\text{percentage error} = & (| \text{ground-truth value} - \text{predicted value} |) \\
                    & /  \text{ground-truth value} * 100  
\end{aligned}
\end{equation}

\subsubsection{Result analysis} The path with the highest risk in each subsystem and combined system is shown in Table~\ref{tab5}. The graphical security model is able to capture the attack path with highest risk and present the sequences of nodes (in a subsystem or across multiple subsystems) that can be potentially compromised by an attacker.

\begin{table}

\caption{The most risky attack paths in subsystems and combined systems}

\centering

\begin{tabular}{|m{2.2cm}|m{5.5cm}|}
\hline
\textbf{Subsystem name} & \textbf{Attack path with highest risk}\\
\hline
Lighting & attacker\textrightarrow motion-sensor\textrightarrow brightness-sensor \textrightarrow light1\textrightarrow occupancy-sensor\textrightarrow light2\textrightarrow light-controller \\
\hline
Audiovisual & attacker\textrightarrow media-player\textrightarrow projector\textrightarrow speaker \\
\hline
Security & attacker \textrightarrow~door-window-alarm-sensor1 \textrightarrow~ door-window-alarm-sensor2 \textrightarrow~burglar-alarm \textrightarrow~surveillance \textrightarrow~entrance-guard \\
\hline
Fire detection & attacker\textrightarrow CO-sensor\textrightarrow fire-alarm \\
\hline
Maintenance & attacker\textrightarrow electrical-current-monitoring-sensor\textrightarrow repair-alarm  \\
\hline
Resource tracking & attacker\textrightarrow wearable-device1\textrightarrow wearable-device2\textrightarrow burglar-alarm  \\
\hline
HVAC & attacker\textrightarrow window-sensor1\textrightarrow window-sensor2\textrightarrow heater \\
\hline
Combined smart building automation & attacker\textrightarrow motion-sensor\textrightarrow brightness-sensor\textrightarrow light1\textrightarrow  light2\textrightarrow light-controller\textrightarrow occupancy-sensor\textrightarrow air-conditioner \\
\hline
Combined security & attacker\textrightarrow door-window-alarm-sensor1\textrightarrow door-window-alarm-sensor2\textrightarrow burglar-alarm\textrightarrow surveillance\textrightarrow entrance-guard \\
\hline
\end{tabular}
\label{tab5}
\end{table}

We show the percentage error of each subsystem and combined system in Table~\ref{tab6}. According to Table~\ref{tab6}, the lowest percentage error of predicting attack probability is 0, while the percentage errors of predicting impact, risk, and base score range from 1.61\% to 33.04\%. Among the subsystems, HVAC has the highest error for risk (i.e., 33.04\%), which is reasonable as it is the most complex subsystem with the most number of devices. The combined smart building automation system contains HVAC subsystem, and thus also has higher errors than the combined security system.

\begin{table}
\caption{Percentage error in predicting probability, impact, risk, and base score for subsystems and combined systems}
\begin{center}
\begin{tabular}{|m{2.2cm}|m{1.2cm}|m{1cm}|m{1cm}|m{0.8cm}|}
\hline
\textbf{Subsystem name} & \textbf{Probability} & \textbf{Impact} & \textbf{Risk} & \textbf{Base score}\\
\hline
Lighting & 0.00\% & 1.48\% & 17.84\% & 11.50\%\\
\hline
Audiovisual & 0.00\% & 2.94\% & 18.62\% & 1.61\% \\
\hline
Security & 0.00\% & 30.41\% & 22.00\% & 11.54\% \\
\hline
Fire detection & 0.00\% & 13.26\% & 8.40\% & 6.72\% \\
\hline
Maintenance system & 0.00\% & 22.83\% & 31.50\% & 15.73\% \\
\hline
Resource tracking & 0.00\% & 13.13\% & 8.97\% & 4.18\% \\
\hline
HVAC & 0.00\% & 2.21\% & 33.04\% & 10.05\% \\
\hline
Combined smart building automation & 0.00\% & 13.60\% & 27.27\% & 17.07\% \\
\hline
Combined security & 0.00\% & 17.73\% & 5.34\% & 11.54\% \\
\hline
\end{tabular}
\label{tab6}
\end{center}
\end{table}

\subsection{Data Visualization}
\begin{figure}
\centerline{\includegraphics[scale=0.70]{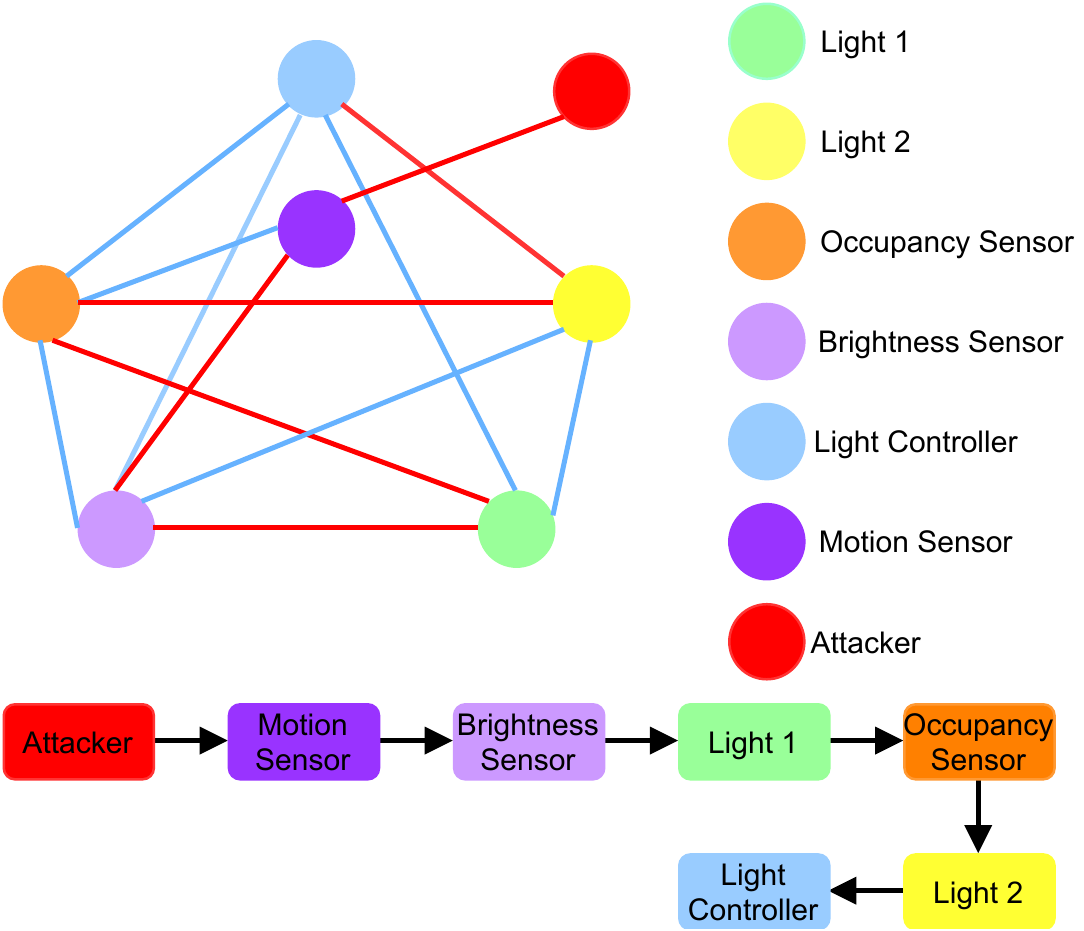}}
\caption{Our web-based visualization of the attack path targeting the Light Controller with the highest risk (highlighted in red color) in the lighting subsystem}
\label{fig:riskiest_path}
\end{figure}

We utilize a network graph by AnyChart \cite{anychart} (i.e., library support for javascript and HTML) to display the network connectivity and the attack path with the highest risk, and merge all the network graphs into a flask project for web presentation. All attack paths in Table~\ref{tab5} can be visualized via the web-based user interface. As an example, we demonstrate the attack path with the highest risk in the lighting subsystem in Fig.~\ref{fig:riskiest_path}, where the sequence of nodes compromised by the attacker along that path are also visualized. Specifically, each node represents a device within the subsystem and has a unique color in the connection diagram. The blue lines indicate the connection between the nodes, while the red line indicates the riskiest attack path of the subsystem.
The bottom part of the figure gives the sequence of the nodes on the most risky path from the attacker's entry point (i.e., Motion Sensor) to the targeted device/node (i.e., Light Controller). 

\section{Discussion} \label{sec:discussion}

In this work, we integrated ML-driven vulnerability assessment and network security assessment for the IoT, and identified a few interesting findings from the analysis results.

\textbf{Over 90\% accurate predictions of security metrics using ML based on NLP:} ML models based on NLP can generate predictions with accuracy of approximately 90\% for all security metrics of each vulnerability. We focus on the professional vulnerability descriptions from the NVD source for both training and testing. Future work can investigate the applicability of the framework to predict user-reported descriptions and descriptions from various databases and sources. Besides formal vulnerability sources like NVD, informal sources such as technical forums like Stack Overflow~\cite{le2021large} can also be explored for security assessment of IoT systems. In addition, the usage of deep learning models~\cite{le2020deep} is worth further exploration, as well as the examination of the degree to which the NLP-based prediction model relies on the professionality and consistency of vulnerability sources.

\textbf{Higher prediction error at network level:} The prediction error is more significant at the network level compared to the errors for individual vulnerability entries. This is expected as the inaccuracy of vulnerability predictions accumulates along with the devices in the attack path. The differences in errors among impact, risk and probability are significant. The predicted probability is identical to the original probability, as the majority of vulnerabilities in the network have a high possibility of being exploited successfully. This demonstrates that the accuracy of the network-level prediction is dependent upon the actual selection of vulnerabilities.

\textbf{Limitations:} A potential issue exists with the resource tracking subsystem. Wearable devices are often movable in a realistic environment. In this work, we only consider the situation that the wearable devices are directly connected to the burglar alarm, and we assume they can maintain the connection when they are moving. Therefore, the topology did not change when we used the HARM as the graphical security model to compute attack paths. However, in a complex network, the topology can change due to moving devices, and the security model needs to be updated whenever the topology changes. To address this issue, an extension to HARM, called temporal HARM can assess dynamically changing networks~\cite{Enoch:Sec_changes2017}. Additionally, the graphical security model has scalability issue when dealing with computationally complex networks, necessitating further optimization of the algorithms for computing attack paths. The impact of this issue can be potentially reduced by dividing the network into multiple sub-networks. 

\section{Conclusion and Future Work} \label{sec:conslusion}
The growing use of the Internet of Things (IoT) exposes these systems to a variety of vulnerabilities due to their hyper-connectivity. However, security assessment based on vulnerability metrics is not fully automated due to manual analysis of vulnerabilities, which is a time-consuming and tedious task.

This paper proposed a framework for automating the security assessment process by employing ML techniques and graphical security modeling.
To evaluate our proposed framework, we developed a proof-of-concept smart building system that simulates real-world IoT systems.
Using data downloaded from the NVD, we filtered IoT device-related vulnerabilities and utilized ML methods to predict Common Vulnerability Scoring System (CVSS) metrics and privileges. The proposed smart building system model and predicted data were used to evaluate the security of IoT systems in terms of probability, risk, and impact using the Graphical Security Model (GSM). The GSM computes all potential attack paths, identifies the most vulnerable path, and visualizes them using a web-based user interface, making it easier for system designers to identify possible vulnerabilities.

In the future, we plan to study various IoT scenarios using the framework and explore the usage of other ML and deep learning techniques to improve the prediction accuracy. Additionally, we will investigate the impact of defences on the change of vulnerability scores to compare the effectiveness of defences.

\bibliographystyle{IEEEtran}
\bibliography{references}

\end{document}